\documentclass[
 aip,
 amsmath,amssymb,
preprint
]{./revtex4-2}

\usepackage{natbib}
\usepackage{graphicx}
\usepackage{dcolumn}
\usepackage{bm}
\usepackage{multirow}
\usepackage{amsmath}
\usepackage{makecell}
\usepackage{tikz}

\usepackage[utf8]{inputenc}
\usepackage[T1]{fontenc}
\usepackage{mathptmx}

\begin{document}


\title{Super-resolution reconstruction of turbulent flow at various Reynolds numbers based on generative adversarial networks}

\author{Mustafa Z. Yousif}
\author{Linqi Yu}

\author{Hee-Chang Lim }
\email[]{Corresponding author, hclim@pusan.ac.kr}
\thanks{}
\affiliation{School of Mechanical Engineering, Pusan National University, 2, Busandaehak-ro 63beon-gil, Geumjeong-gu, Busan, 46241, Rep. of KOREA}

\date{\today}

\begin{abstract}
This study presents a deep learning-based framework to reconstruct high-resolution turbulent velocity fields from extremely low-resolution data at various Reynolds numbers using the concept of generative adversarial networks (GANs). A multiscale enhanced super-resolution generative adversarial network (MS-ESRGAN) is applied as a model to reconstruct the high-resolution velocity fields, and direct numerical simulation (DNS) data of turbulent channel flow with large longitudinal ribs at various Reynolds numbers are used to evaluate the performance of the model. The model is found to have the capacity to accurately reproduce high-resolution velocity fields from data at two different low-resolution levels in terms of the quantities of velocity fields and turbulent statistics. The results further reveal that the model is able to reconstruct velocity fields at Reynolds numbers that are not used in the training process.

\end{abstract}

\maketitle

\section{Introduction}\label{sec:introduction}
Turbulence has a chaotic nature with multiple spatio-temporal flow scales, making experimental measurements and numerical simulations unpredictable and sophisticated and requiring a considerable cost to describe the flow structure effectively. At the same time, an abundance of high-fidelity data can be generated using state-of-the-art particle image velocimetry (PIV) and direct numerical simulation (DNS). This opens the door to the discovery of new data-driven methods able to make use of the data and the construction of models that can tackle different problems in the research area of turbulent flows. Machine learning has demonstrated a strong ability to tackle highly non-linear problems in various fields. Further, in recent years, many deep learning algorithms have been applied practically in the field of fluid dynamics \cite{Bruntonetal2020,  Kutz2017}. Deep learning is a subset of machine learning in which deep neural networks are used for classification, prediction and feature extraction \cite{LeCunetal2015}.
Various deep learning-based approaches have been recently applied to deal with various turbulence problems, including turbulence modelling \cite{Duraisamyetal2019, Gamahara&Hattori2017, Lingetal2016, Wangetal2018}, turbulent flow prediction \cite{Lee&You2019, Srinivasanetal2019} and turbulent flow control \cite{Fanetal2020, Rabaultetal2019}. In terms of the super-resolution reconstruction of turbulent flow fields, deep learning-based high-resolution image reconstruction has exhibited great potential for practical application in the reconstruction of high-fidelity turbulent flows. This can be attributed to the recent development in supervised and unsupervised deep learning algorithms that significantly outperform traditional handcrafted super-resolution techniques, such as bicubic interpolation \cite{Bashiretal2021, Keys1981}.
Motivated by the remarkable ability of convolutional neural networks (CNNs) to tackle the spatial mapping of data, Fukami {\it et al.} \cite{Fukami2019, Fukami2021} proposed models based on a CNN for the spatial and spatio-temporal super-resolution reconstruction of turbulent flows. Their results revealed satisfactory outcomes in the reconstruction of velocity and vorticity fields from extremely low-resolution data and a good prediction of the temporal evolution for the intervals the model was trained for. \par

Liu {\it et al.} \cite{Liuetal2020} applied a CNN-based model to reconstruct high-resolution turbulent flow data from low-resolution data with multiple temporal paths and reported that their model had better reconstruction performance compared with static CNN-based models. Kim {\it et al.} \cite{Kimetal2021} applied a cycle-consistent generative adversarial network (CycleGAN) \cite{Zhuetal2017} to reconstruct flow fields from low-resolution DNS and LES data as an unsupervised deep learning technique to reconstruct turbulence with high resolution using unpaired training data. Their results revealed that the method had better reconstruction accuracy compared with a bicubic interpolation and CNN-based model. \par

Cai {\it et al.} \cite{Caietal2019} presented a model based on CNN for estimating high-resolution velocity fields from PIV measurements and reported that their model had better performance than conventional cross-correlation algorithms. Further, Morimoto {\it et al.} \cite{Morimotoetal2021} applied a CNN-based model to estimate velocity fields using PIV measurements of flow around a square cylinder with missing regions. Their results revealed a relatively good estimation of the missing regions in the velocity fields. Deng {\it et al.} \cite{Dengetal2019} applied a super-resolution generative adversarial network (SRGAN) \cite{Ledigetal2017} and enhanced SRGAN (ESRGAN) \cite{Wangetal2018} to reconstruct high-resolution velocity fields using PIV measurements of flow around a cylinder. Their results demonstrated good agreement with the ground truth data, and ESRGAN was found to have better reconstruction capability compared with SRGAN. Recently, Yousif {\it et al.} \cite{Yousifetal2021} proposed multiscale ESRGAN (MS-ESRGAN) with a physics-based loss function to reconstruct high-fidelity turbulent flow fields from spatially limited data. They reported that the model has a strong ability to reconstruct high-resolution turbulence even from extremely coarse data. \par

In all the aforementioned studies, the models were trained to reconstruct high-resolution flow fields at a specific Reynolds number. However, most turbulent flow problems, in which the flow parameters can be changed spatially and temporally, result in various Reynolds numbers. \par

In this study, we consider the reconstruction of high-resolution turbulent velocity fields at certain salient Reynolds numbers. In particular, we apply MS-ESRGAN to reconstruct high-resolution velocity fields at different Reynolds numbers from extremely low-resolution data. To evaluate the performance of MS-ESGAN, DNS calculations of turbulent channel flow with large longitudinal ribs are performed at different Reynolds numbers. \par

The remainder of this paper is organised as follows. Section 2 explains the methodology for reconstructing high-resolution flow fields using the proposed deep learning model. Section 3 describes the generation of the training data using DNS. Section 4 discusses the testing results obtained using the proposed model. Finally, Section 5 presents the conclusions of the study. \par

\section{Methodology}

Goodfellow {\it et al.} \cite{Goodfellowetal2014} first introduced the concept of  the generative adversarial networks (GANs). Different types of GANs have since been proposed to tackle various image transformation and super-resolution problems \cite{Ledigetal2017, Mirza&Osindero2014, Wangetal2018, Zhuetal2017}. In GAN, two adversarial neural networks, called the generator ($G$) and the discriminator ($D$), compete with each other. In particular, $G$ generates fake images similar to real images, and $D$ tries to distinguish the fake ones from the real ones. The goal of the training process is to train $G$ to generate fake images that are difficult for $D$ to distinguish. This process can be expressed as a min-max two-player game with a value function $V(D, G)$ such that:

\begin{equation} \label{eqn:eq1}
\begin{split}
\substack{min\\G} ~\substack{max\\D} ~V(D,G) = \mathbb{E}_{x_r \sim P_{data}(x_r)} [ {\rm log} D(x_r )] + \mathbb{E}_{z \sim P_z(z) } [ {\rm log} (1-D(G(z)))],
\end{split}
\end{equation}

\noindent where $x_r$ is the image from the ground truth data and $P_{data}(x_r )$ is the distribution of the real image. $\mathbb{E}$ represents the operation used to calculate the average of all the data in the training mini-batch. In the term second to the right in Eq.~\ref{eqn:eq1}, $z$ is a random vector used as an input to $G$, and $D(x_r)$ represents the probability that the image is real and not generated by the generator. $G(z)$ is the output from $G$ and is expected to generate an image that is similar to the real image, such that the value of $D(G(z))$ becomes as large as possible. At the same time, $D$ attempts to increase the value of $D(x_r )$ and decrease the value of $D(G(z))$. Thus, during the training process, $G$ is trained to minimise $V(D,G)$, and $D$ is trained to maximise $V(D,G)$. After successful training, $G$ is expected to produce an image with a distribution similar to the real image that $D$ cannot distinguish from the real image.

\begin{figure}
\centering 
\includegraphics[angle=0, trim=0 0 0 0, width=0.9\textwidth]{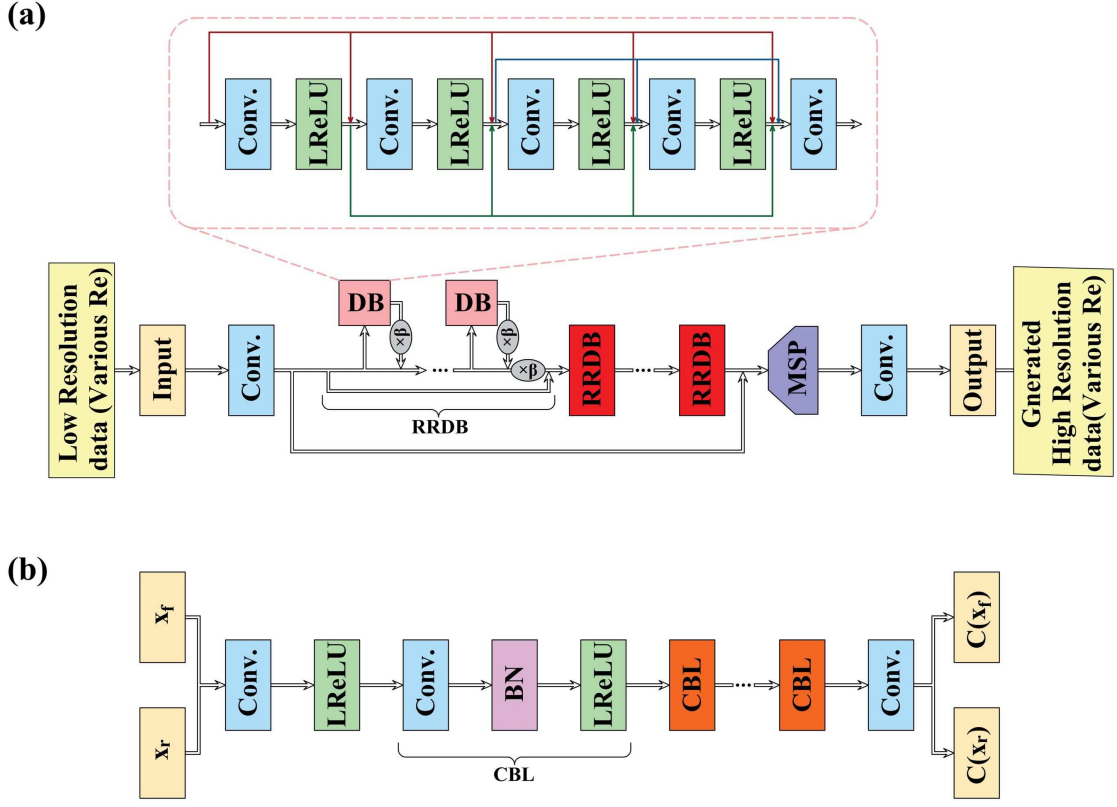}
\caption[]{MS-ESRGAN architecture: (a) the generator ($\beta =0.2$  is the residual scaling parameter), and (b) the discriminator}
\label{fig:1-ESRGAN}
\end{figure}

This study uses a deep learning framework based on ESRGAN \cite{Wangetal2018} to reconstruct high-resolution flow fields from extremely low-resolution data. We applied the newly developed MS-ESRGAN model \cite{Yousifetal2021} to reconstruct the high-resolution velocity fields at different Reynolds numbers. Figure~\ref{fig:1-ESRGAN}(a) and (b) shows the basic architectures of $G$ and $D$ in the MS-ESRGAN, respectively. $G$ consists of a deep convolution neural network represented by residuals in residual dense blocks (RRDBs) and multiscale part (MSP) . The low-resolution input image is first fed to $G$ and passed through a convolution layer and then through a series of RRDBs. MSP, consisting of three parallel convolutional sub-models with different kernel sizes, is applied to the data features that the RRDBs extract, as explained in Table~\ref{tab:MSP}. Finally, the outputs of the three branches are summed and passed through a final convolutional layer to generate a high-resolution fake image $(x_f)$. The fake and real images are fed to $D$ and passed through a series of convolutional, batch normalisation (BN) and leaky rectified linear unit (ReLU) layers. The data are then passed through a final convolutional layer. The non-transformed discriminator outputs using the real and fake images $C(x_r)$ and $C(x_f)$ are used to calculate the relativistic average discriminator value ($D_{Ra}$) \cite{Jolicoeur-Martineau2018} as follows:

\begin{equation} \label{eqn:eq2}
D_{Ra} (x_r , x_f ) = \sigma (C \left(x_r ) \right) - \mathbb{E}_{x_f} \left[C ( x_f) \right],
\end{equation}

\begin{equation} \label{eqn:eq3}
D_{Ra} (x_f , x_r ) = \sigma (C \left(x_f ) \right) - \mathbb{E}_{x_r} \left[C ( x_r) \right],
\end{equation}

\noindent where $\sigma$ is the sigmoid function. In Eqs.~\ref{eqn:eq2} and ~\ref{eqn:eq3}, $D_{Ra}$ represents the probability that the output from $D$ using the real image is more realistic than the output using the fake image. \par

The discriminator loss is then defined as:

\begin{equation} \label{eqn:eq4}
L_D^{Ra} = -\mathbb{E}_{x_r} \left[ {\rm log} (D_{Ra} (x_r , x_f )) \right] - \mathbb{E}_{x_f} \left[ {\rm log} (1 - D_{Ra} (x_f , x_r )) \right].
\end{equation}

The adversarial loss of the generator can be expressed in a symmetrical form as:

\begin{equation} \label{eqn:eq5}
L_G^{Ra} = -\mathbb{E}_{x_r} \left[ {\rm log} (1 - D_{Ra} (x_r , x_f )) \right] - \mathbb{E}_{x_f} \left[ {\rm log} (D_{Ra} (x_f , x_r )) \right].
\end{equation}

The combined loss function of the generator  is defined as:

\begin{equation} \label{eqn:eq6}
\mathcal{L}_G = \lambda_1L_G^{Ra} + \lambda_2 L_{pixel} + \lambda_3 L_{perceptual},     
\end{equation}

\noindent where $L_{pixel}$ is the error calculated based on the pixel difference between the generated data and the data obtained from the ground truth data. $L_{perceptual}$ represents the difference in the features extracted from the real and fake data. The pre-trained CNN VGG-19 \cite{Simonyan&Zisserman2015} is employed to extract the features by using the output of three different layers\cite{Yousifetal2021}. Here, $\lambda_1$, $\lambda_2$ and $\lambda_3$ are the coefficients used to balance the loss terms whose values are empirically set to be 2, 2,000 and 4, respectively. In this study, the mean squared error (MSE) is used to calculate $L_{pixel}$ and $L_{perceptual}$. The adaptive moment estimation (ADAM) optimisation algorithm \cite{Kingma&Ba2017} is applied to update the model weights. The training data are divided into mini-batches, and the size of each mini-batch is set to 16.  \par

\begin{table} 
\begin{center}
~\caption{MSP architecture}
\scalebox{0.9}{
\begin{tabular}{ p{5cm} p{5cm} p{5cm} }
\hline\hline
 $1^{st}$ branch 		& $2^{nd}$ branch 		&  $3^{rd}$ branch 	\\ \hline\hline
 Conv.(3, 3) 			&  Conv.(5, 5) 			&  Conv.(7, 7) 		\\ 
 UpSampling(2, 2)		&  UpSampling(2, 2)		&  UpSampling(2, 2)	 \\  
 Conv.(3, 3) 			&  Conv.(5, 5) 			&  Conv.(7, 7) 		\\ 
 LeakyReLU			&  LeakyReLU			&  LeakyReLU		\\  
 UpSampling(2, 2)		&  UpSampling(2, 2)		&  UpSampling(2, 2)	 \\ 
 Conv.(3, 3) 			&  Conv.(5, 5) 			&  Conv.(7, 7) 		\\ 
 LeakyReLU			&  LeakyReLU			&  LeakyReLU		\\  
 UpSampling(2, 2)		&  UpSampling(2, 2)		&  UpSampling(2, 2)	 \\  
 Conv.(3, 3) 			&  Conv.(5, 5) 			&  Conv.(7, 7) 		\\ 
 LeakyReLU			&  LeakyReLU			&  LeakyReLU		\\  \hline
 \multicolumn{3}{c} {Add ($1^{st}$ branch,  $2^{nd}$ branch, $3^{rd}$ branch) } \\ \hline \hline
\end{tabular}} \label{tab:MSP}
\end{center}
\end{table}

\section{Training data generation}

\begin{figure}
\centering 
\includegraphics[angle=0, trim=0 0 0 0, width=0.8\textwidth]{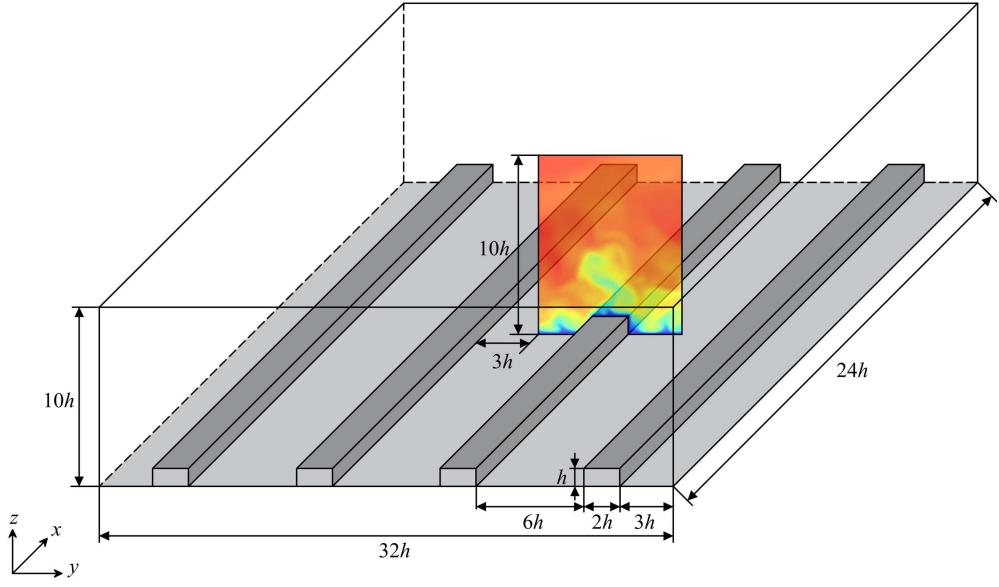}
\caption[]{Schematic of the computational domain} 
\label{fig:2-Domain}
\end{figure}

In this study, a fully developed turbulent channel flow with large longitudinal ribs is used as a test case. This case has been widely studied experimentally and numerically for a wide range of Reynolds numbers \cite{Castroetal2021, Hwang&Lee2018, Vanderweletal2019, Vanderwel&Ganapathisubramani2015, Zampironetal2020}, making it suitable for assessing our model. To generate the data, the DNS calculation is performed. Figure~\ref{fig:2-Domain} shows the schematic of the computational domain used in this study. The domain contained four large longitudinal ribs along the streamwise direction ($x$) and equally distributed in the spanwise direction ($y$). The domain dimensions are $24h$, $32h$ and $10h$ in the $x, y$ and the ground-normal ($z$) directions, respectively. Here, $h$ is the height of each rib, which is set as 1.  The size of each rib is $24h \times 2h \times h$. A section normal to the streamwise direction that contains one of the ribs is used to reconstruct the high-resolution velocity fields. \par

The momentum equation for an incompressible viscous fluid can be expressed as:

\begin{equation}\label{eqn:eq7}
\frac{\partial {\bf u}}{\partial t} + {\bf u}\cdot \nabla {\bf u} = - \frac{1}{\rho} \nabla p + {\it \nu} \nabla^2 {\bf u},
\end{equation}

\noindent where ${\bf u}$ is the velocity of the fluid, $\rho$ is the density, $p$ is the pressure and $\nu$ is the kinematic viscosity. The continuity equation of the fluid is defined as:

\begin{equation}\label{eqn:eq8}
\nabla \cdot {\bf u} = 0.
\end{equation}

A version of the in-house finite-difference code CGLES, originally written by Thomas and Williams \cite{Thomas&Williams1997}, is used to perform the DNS calculation. The friction Reynolds numbers used for training in this study ($Re_\tau = u_\tau H / \nu $) are set to 292, 450 and 850, respectively, where $u_\tau$ is the friction velocity and $H$ is the channel height. Cartesian uniform meshes are used in all directions, and the second-order central differencing scheme is applied to all spatial derivatives. The Adams-Bashforth scheme is employed for time advancement using the pressure projection method. The continuity is forced implicitly by solving the Poisson equation for pressure using a parallel multi-grid method. A no-slip condition is applied at the ground and the walls of the ribs, and the periodic boundary condition is applied in the streamwise and spanwise directions, while a free-slip condition is applied to the top of the domain. The flow is driven by applying a pressure gradient. \par
To avoid any interpolation of the training data at the three Reynolds numbers used in this study, the mesh size is fixed to $768 \times 1024 \times 320$, with element dimension $\Delta x^+ = \Delta y^+ = \Delta z^+  \approx 2.7$. The time step $\Delta t^+ $ is set as 0.21, where the superscript $+$ indicates that the quantity is made dimensionless using the wall units $\nu$ and $u_\tau$. The grid size is carefully chosen to ensure it is sufficient to solve all the turbulent flow scales of the case with the highest $Re_\tau$, i.e. 850. This could be achieved by estimating the smallest Kolmogorov length scale $\eta$ in the flow. By assuming that the energy production roughly balances the dissipation, we can write the following for the latter: $\varepsilon = - \overline{u' w'} dU / dz $, where $\overline {u' w'} $ is Reynolds shear stress; $u'$ and $w'$ are the fluctuations of the streamwise and ground-normal velocity components, respectively and $U$ is the mean streamwise velocity. It is found that at $z \approx 2h$, the shear stress has a maximum value of approximately $0.7 u_\tau^2$, so that  $\eta \approx$ $1.3 u_\tau^3 ⁄ h$. Hence, $\eta \approx 0.0056 h = 0.18 \Delta$, leading to $\Delta \approx 0.6 \eta$. Moin \& Mahesh \cite{Moin&Mahesh1998} explain that the smallest length scale that must be resolved is typically much larger than the Kolmogorov scale ($\sim 10 \eta$ ). Based on this fact, we can ensure that the present simulation setup captures the entire dissipation spectrum for all the Reynolds numbers used in this study. At the same time, the maximum Courant-Friedrichs-Lewy number (CFL) is maintained at $<0.3$ to ensure the stability of the simulations. \par

The training data are obtained from each simulation with 10,000 snapshots from a single ($y-z$) plane. The size of the plane is fixed as $y/h \times z/h = 8 \times 10$, which is equivalent to a grid size of $256 \times 320$. The interval between the collected snapshots of the flow fields is ten times the time step used in the simulations. The low-resolution training data are obtained by downsampling the data from the DNS using max pooling. Two downsampling levels are used in this study, $\times 8$ and $\times 16$, which are equivalent to sections of sizes of $32 \times 40$ and $16 \times 20$, respectively. To prepare the data for training, they are normalised using the min–max normalisation function to produce values between 0 and 1.

\section{Results and discussion}

\subsection{Velocity fields reconstruction (Reynolds numbers are used in the training process)}

In this section, the capacity of the model to reconstruct high-resolution velocity fields at the three Reynolds numbers used in the training process ($Re_\tau$  = 292, 550 and 850) from the downsampled data at two low-resolution levels ($\times 8$ and $\times 16$) is thoroughly examined and discussed. It is worth noting that all of the reconstructed velocity fields are based on the test data that are not used in the training  process. \par

\begin{figure}
\centering 
\includegraphics[angle=0, trim=0 0 0 0, width=0.8\textwidth]{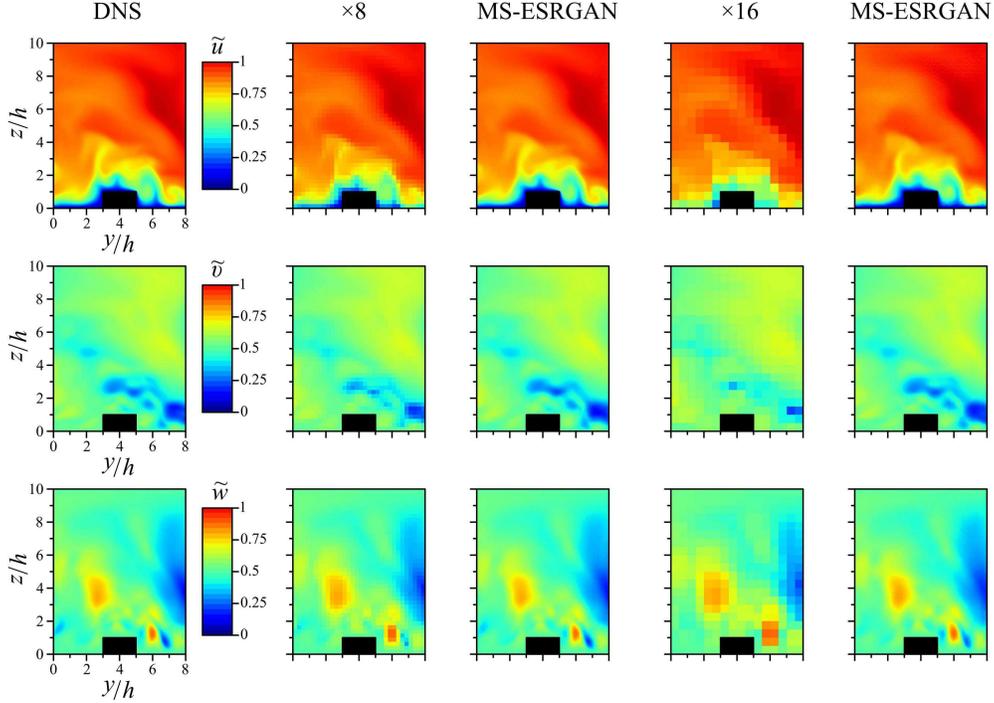}
\caption[]{Reconstructed instantaneous velocity fields at $Re_\tau  = 292$} 
\label{fig:3-RecVel1}
\end{figure}

\begin{figure}
\centering 
\includegraphics[angle=0, trim=0 0 0 0, width=0.8\textwidth]{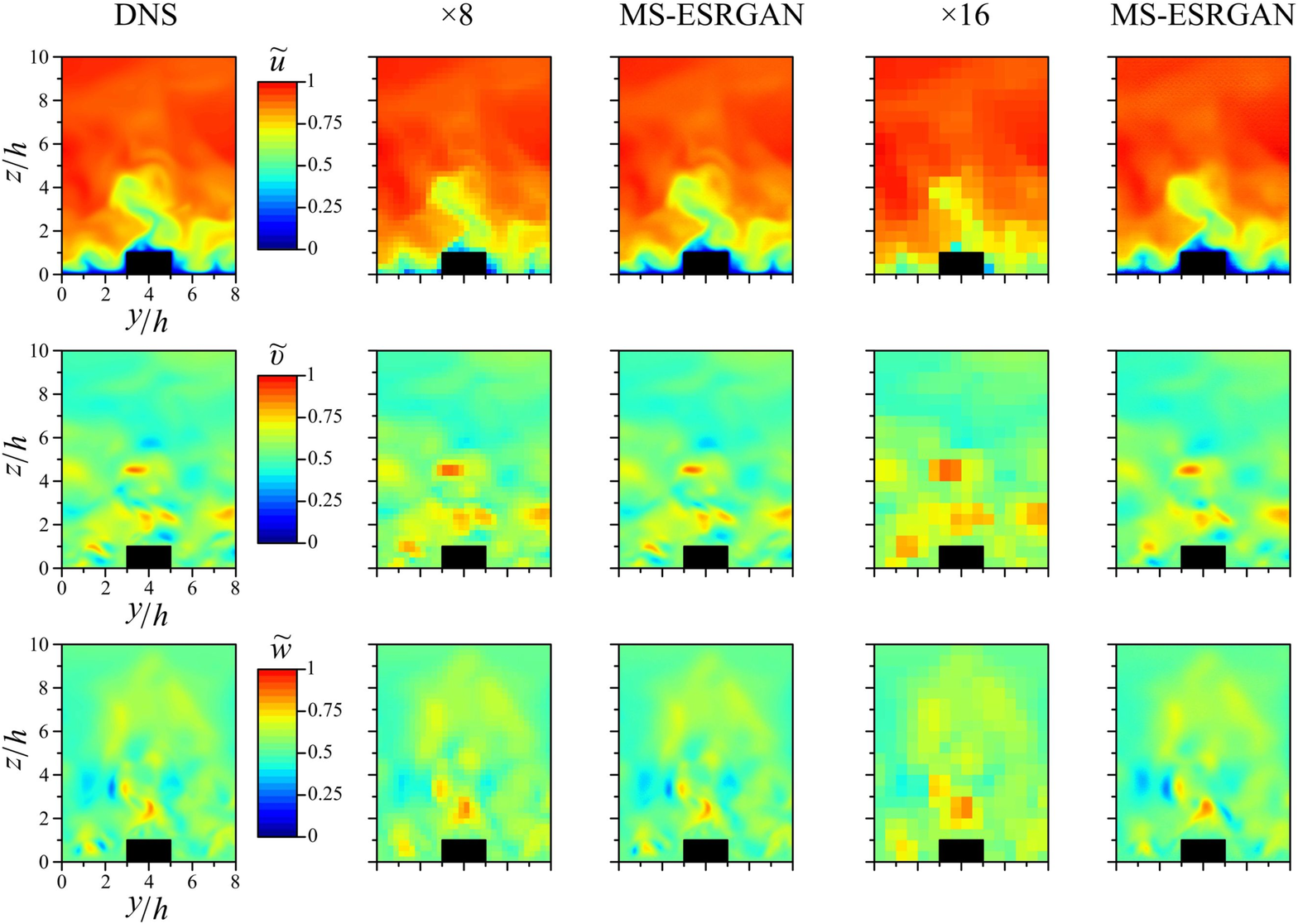}
\caption[]{Reconstructed instantaneous velocity fields at $Re_\tau  = 550$} 
\label{fig:4-RecVel2}
\end{figure}

\begin{figure}
\centering 
\includegraphics[angle=0, trim=0 0 0 0, width=0.8\textwidth]{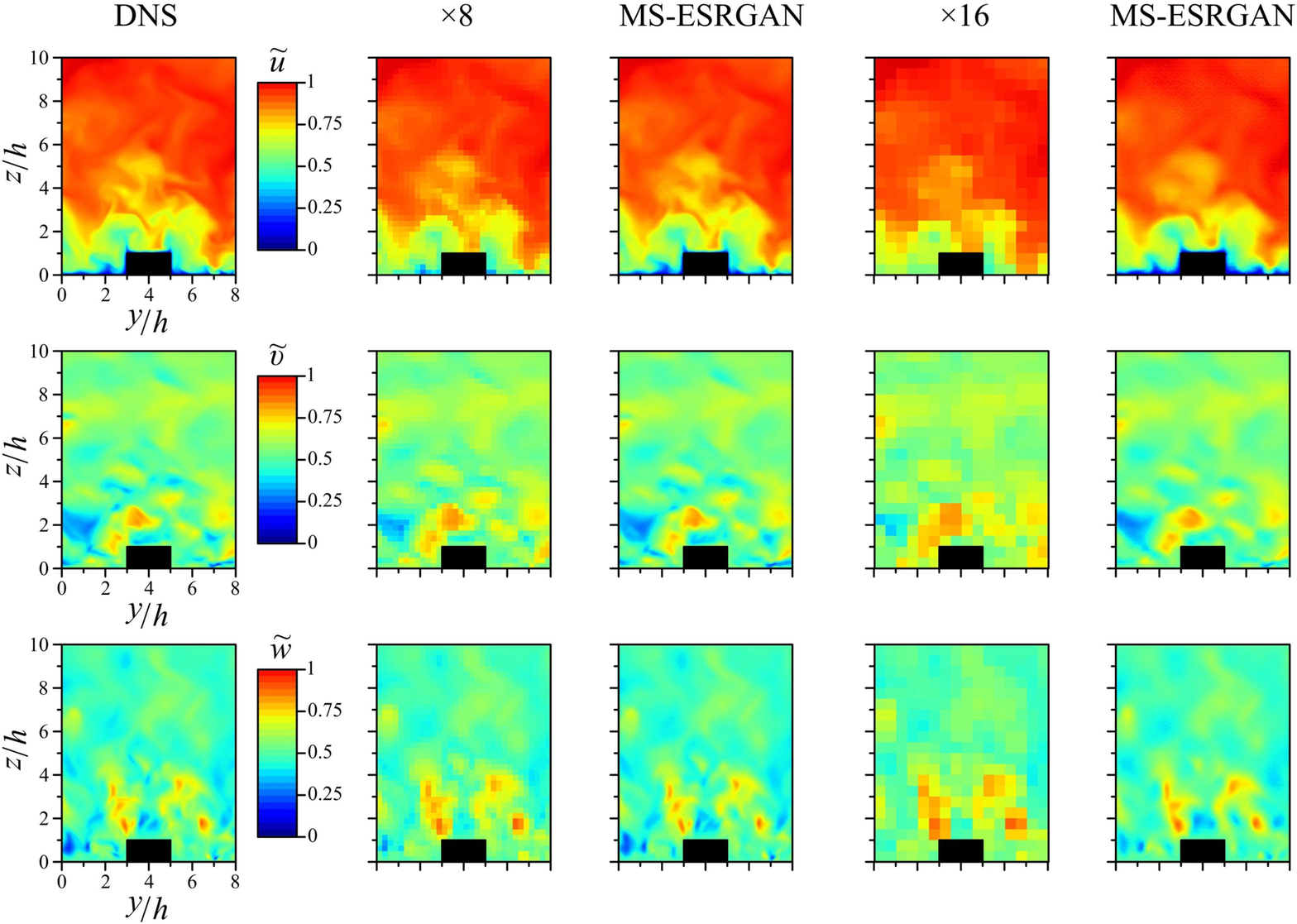}
\caption[]{Reconstructed instantaneous velocity fields at $Re_\tau  = 850$} 
\label{fig:5-RecVel3}
\end{figure}

Figures~\ref{fig:3-RecVel1}-\ref{fig:5-RecVel3} show the reconstructed instantaneous fields of the streamwise velocity ($\tilde{u}$), spanwise velocity ($\tilde{v}$) and ground-normal velocity ($\tilde{w}$) at the three Reynolds numbers and two resolution levels used in this study. Here, $\sim$ indicates that the quantity is normalised using the min–max normalisation function. The reconstructed velocity fields closely match the DNS data, as indicated in the figures. When the $\times 8$ resolution level is employed, there is no discernible change in the reconstructed velocity fields. Nonetheless, there is a slight under-prediction in several velocity fields regions when the $\times 16$ resolution level is used. In addition, increasing the Reynolds number is found to have no influence on the correctness of the reconstruction, with the instantaneous velocity fields being successfully reconstructed at all three Reynolds numbers. \par

\begin{figure}
\centering 
\includegraphics[angle=0, trim=0 0 0 0, width=0.8\textwidth]{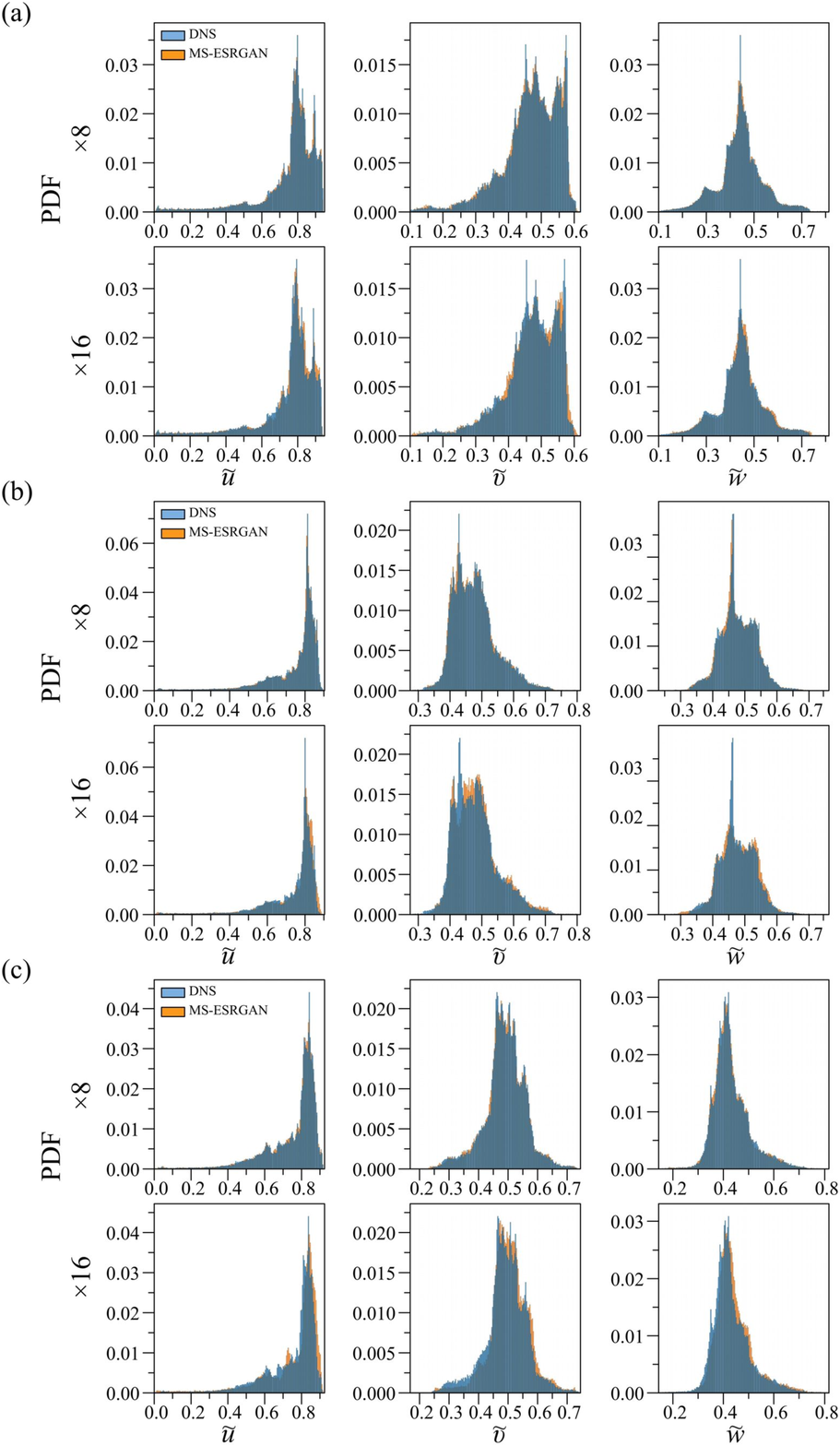}
\caption[]{Probability density functions of the reconstructed velocity components at (a) $Re_\tau  = 292$, (b) $Re_\tau  = 550$ and (c) $Re_\tau  = 850$} 
\label{fig:6-RecPDF}
\end{figure}

Figure~\ref{fig:6-RecPDF} shows the probability density function (PDF) of the reconstructed velocity components. The figure reveals a reasonable level of agreement between the PDF of the reconstructed velocity fields and that obtained from the DNS data, even for the highest Reynolds number and the $\times 16$ resolution level, demonstrating the ability of the model to reconstruct high-resolution velocity fields for all three Reynolds numbers and the two low-resolution levels. \par

\begin{figure}
\centering 
\includegraphics[angle=0, trim=0 0 0 0, width=1.1\textwidth]{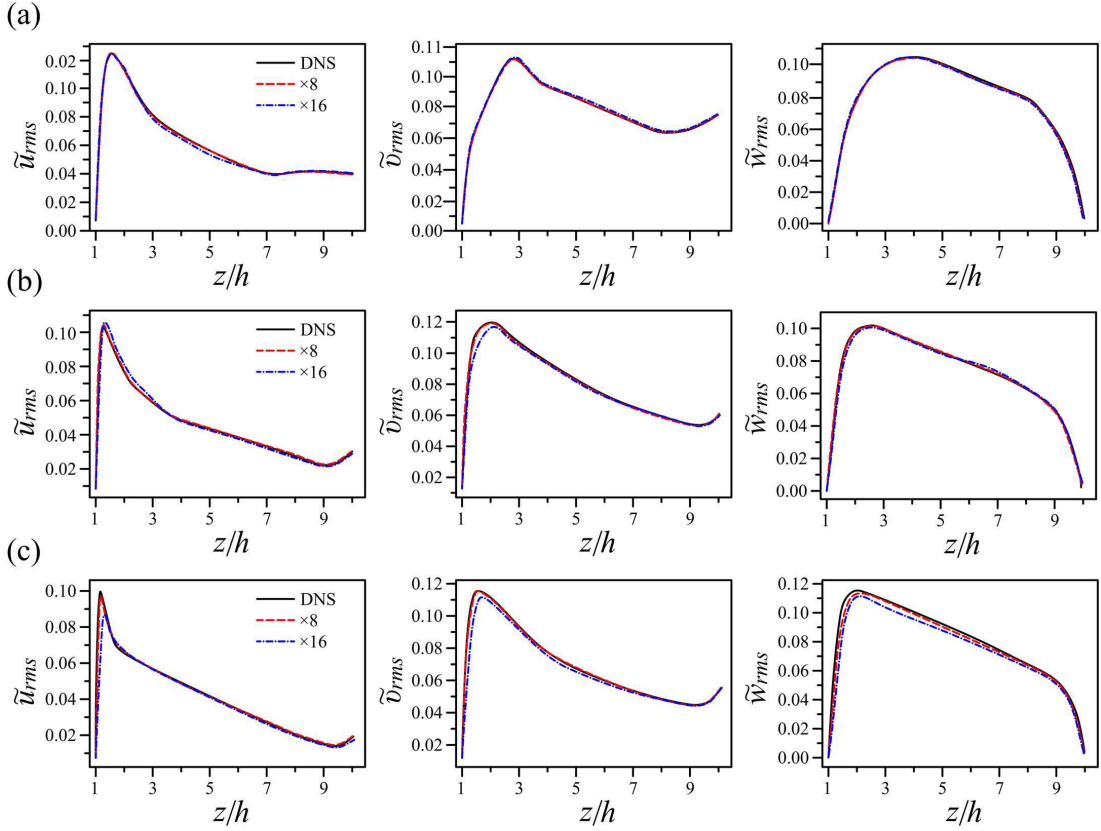}
\caption[]{RMS profiles of the reconstructed velocity components at (a) $Re_\tau  = 292$, (b) $Re_\tau  = 550$ and (c) $Re_\tau  = 850$} 
\label{fig:7-RecRMS}
\end{figure}

Figure~\ref{fig:7-RecRMS} shows the root mean square (RMS) profiles of the reconstructed velocity components ($\tilde{u}_{rms}$, $\tilde{v}_{rms}$ and $\tilde{w}_{rms}$) averaged over 100,000 snapshots and along the upper wall of the rib. It can be observed from the figure that the RMS values of all the velocity components are in fair agreement with the DNS results when the $\times 8$ data are used and for all three Reynolds numbers. However, a slight deviation is observed for the case in which the $\times 16$ data are used. This can be attributed to the limited information on the high-frequency fluctuations of the velocity components when extremely low-resolution data are used. \par

\begin{figure}
\centering 
\includegraphics[angle=0, trim=0 0 0 0, width=0.9\textwidth]{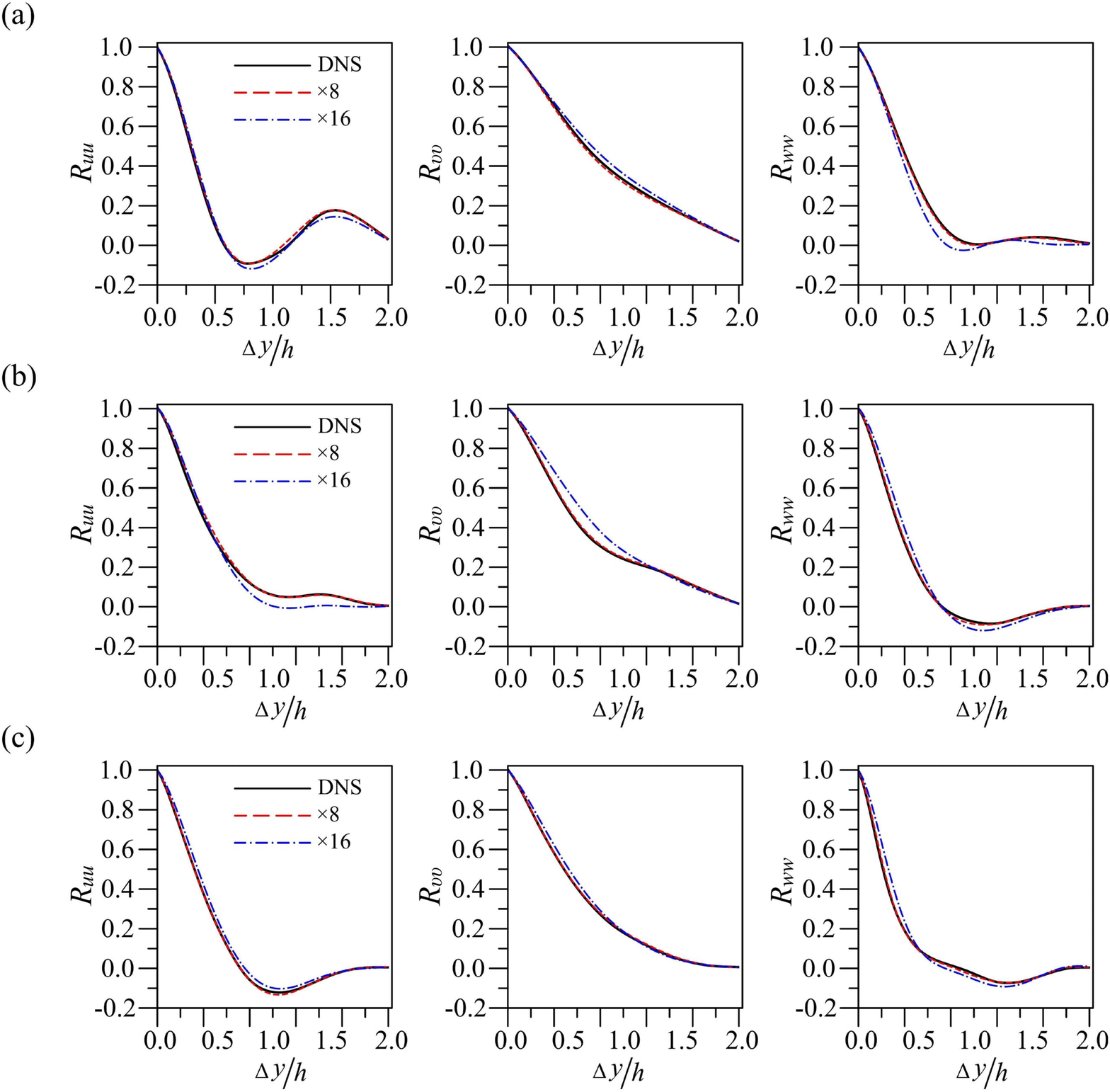}
\caption[]{Two-point correlations of the reconstructed velocity components at (a) $Re_\tau  = 292$, (b) $Re_\tau  = 550$ and (c) $Re_\tau  = 850$} 
\label{fig:8-Rec2Correl}
\end{figure}

To estimate how exact it is possible to make the spatial correlations of the reconstructed velocity components, the spanwise two-point correlation ($R_{ii} (\Delta y )$) of each of the reconstructed velocity components is calculated along the upper wall of the rib at $z = 2h$. Here, $i$ represents the velocity component. As shown in Figure~\ref{fig:8-Rec2Correl}, the correlations are generally in good agreement with the results from the DNS, with no reduction in the accuracy caused by increasing the Reynolds number. Nevertheless, an offset is observed when extremely low-resolution data are used, which can be attributed to the insufficient spatial information on the velocity components.
All the previous results reveal that the model used in this study is able to successfully reproduce high-resolution velocity fields with a commendable accuracy not affected by increasing the Reynolds number. Nevertheless, they also demonstrate that the reconstruction accuracy is slightly affected when extremely low-resolution data are used as input to the model. \par

\subsection{Velocity fields reconstruction (Reynolds numbers are not used in the training process)}

The capability of the model is further examined using data of velocity fileds at Reynolds numbers that are not used in the training process. Here, two Reynolds numbers are used, $Re_\tau $ = 400 and 600 at two different low-resolution levels ( $\times 8$ and $\times 16$). The test is used to investigate the ability of the model to interpolate the reconstructed velocity fields, since the used Reynolds numbers are in the range of those the model is trained with. \par

\begin{figure}
\centering 
\includegraphics[angle=0, trim=0 0 0 0, width=0.8\textwidth]{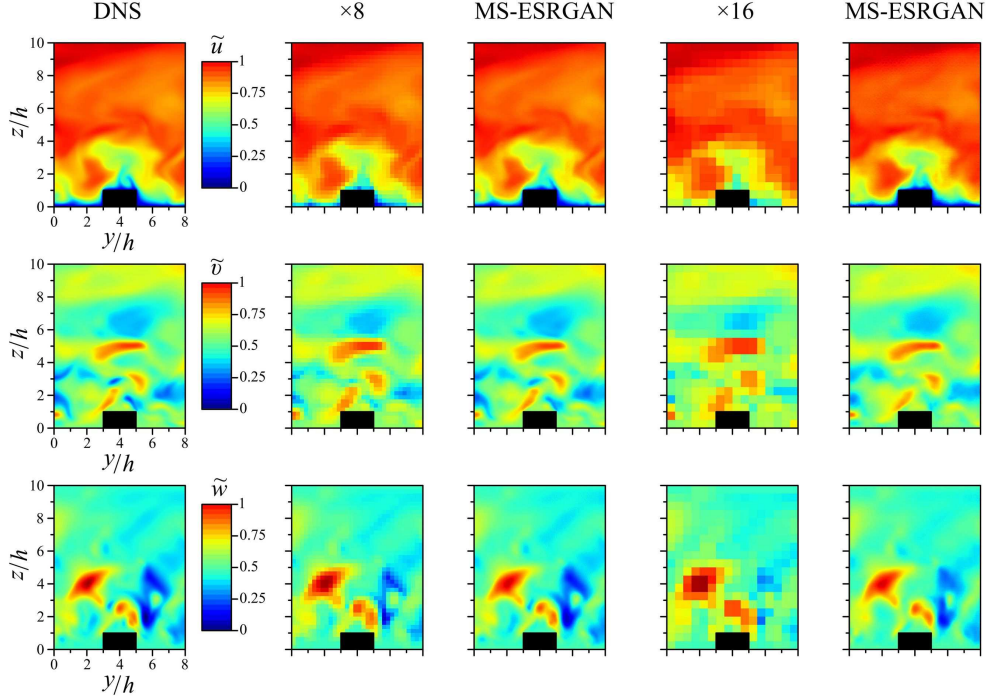}
\caption[]{Reconstructed instantaneous velocity fields at $Re_\tau = 400$} 
\label{fig:9-RecVel-400}
\end{figure}

\begin{figure}
\centering 
\includegraphics[angle=0, trim=0 0 0 0, width=0.8\textwidth]{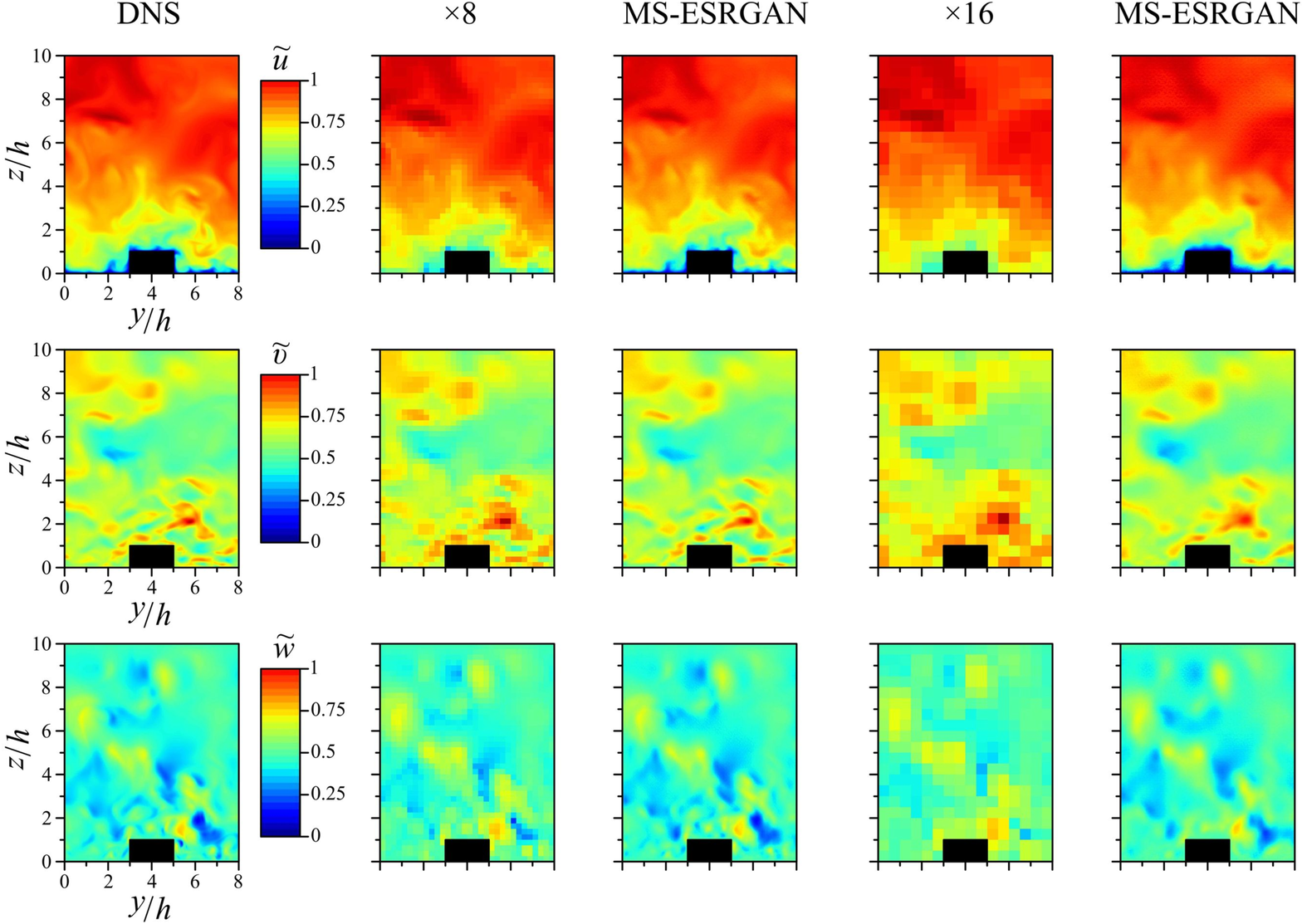}
\caption[]{Reconstructed instantaneous velocity fields at $Re_\tau = 600$} 
\label{fig:10-RecVel-600}
\end{figure}

As shown in Figures~\ref{fig:9-RecVel-400} and \ref{fig:10-RecVel-600}, the reconstructed instantaneous velocity fields exhibit acceptable agreement with the DNS data for both tested Reynolds numbers. The reconstructed velocity fields also exhibit a similar level of precision when the training Reynolds numbers are used. In addition, increasing the Reynolds number is found to have no influence on the accuracy of the reconstruction. \par

\begin{figure}
\centering 
\includegraphics[angle=0, trim=0 0 0 0, width=0.8\textwidth]{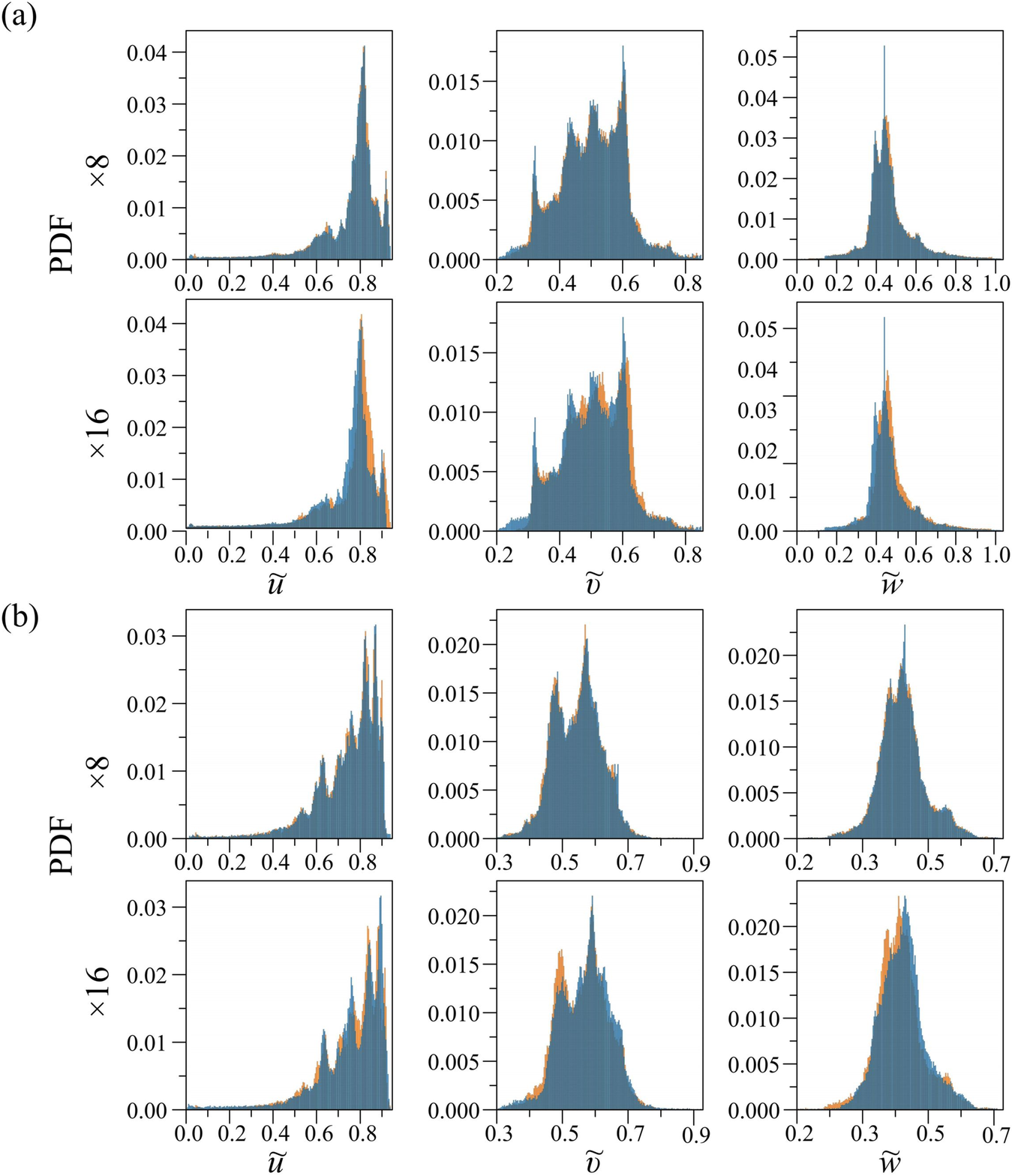}
\caption[]{Probability density functions of the reconstructed velocity components at (a) $Re_\tau = 400$ and (b) $Re_\tau = 600$} 
\label{fig:11-RecPDF-400-600}
\end{figure}

As shown in Figure~\ref{fig:11-RecPDF-400-600}, the PDF of the reconstructed velocity components shows no significant difference from that obtained from the DNS data. However, the difference is more noticeable when the $\times 16$ low-resolution level is used than when the training Reynolds numbers are used. As mentioned in the previous section, this can be attributed to the effect of having limited spatial information, which influenced the mapping ability of the model, resulting in less-accurate interpolation in the regions of the small-scale eddies. \par

\begin{figure}
\centering 
\includegraphics[angle=0, trim=0 0 0 0, width=1.0\textwidth]{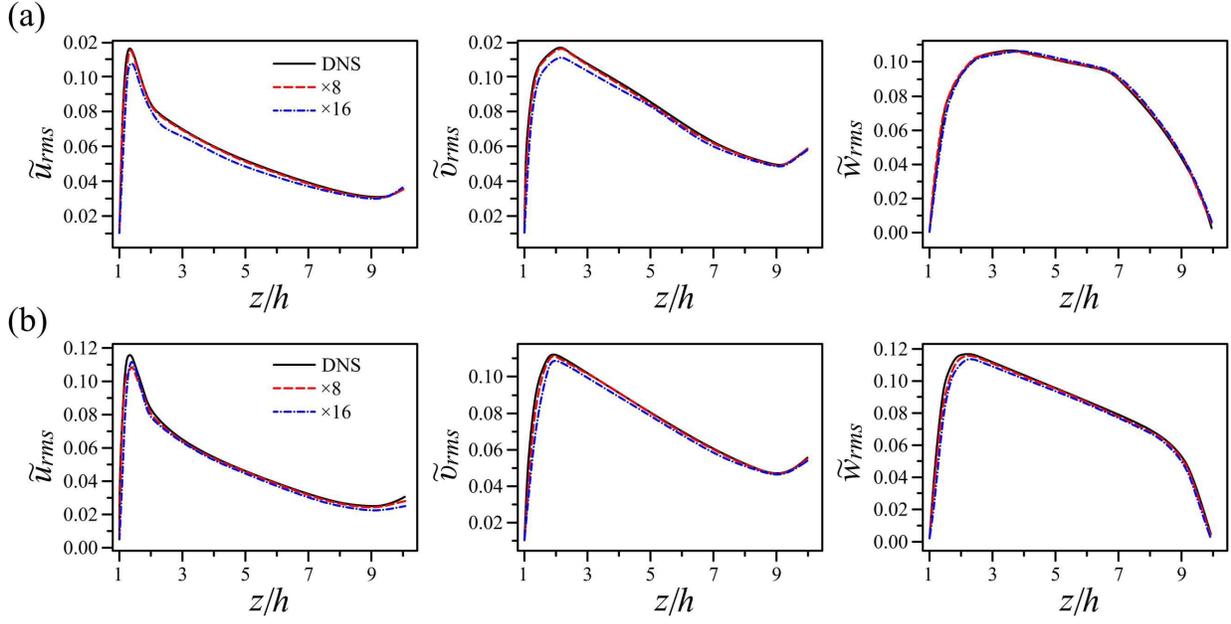}
\caption[]{RMS profiles of the reconstructed velocity components at (a) $Re_\tau = 400$ and (b) $Re_\tau = 600$} 
\label{fig:12-RecRMS-292-550}
\end{figure}

The RMS profiles of the reconstructed velocity fields as plotted in Figure~\ref{fig:12-RecRMS-292-550} show no noticeable difference with the DNS results when the $\times 8$ data are used. However, similar to the case in which the training Reynolds numbers are used, a slight under- and over-prediction can be seen in the RMS values in the case when $\times 16$ data are used. \par

\begin{figure}
\centering 
\includegraphics[angle=0, trim=0 0 0 0, width=1.0\textwidth]{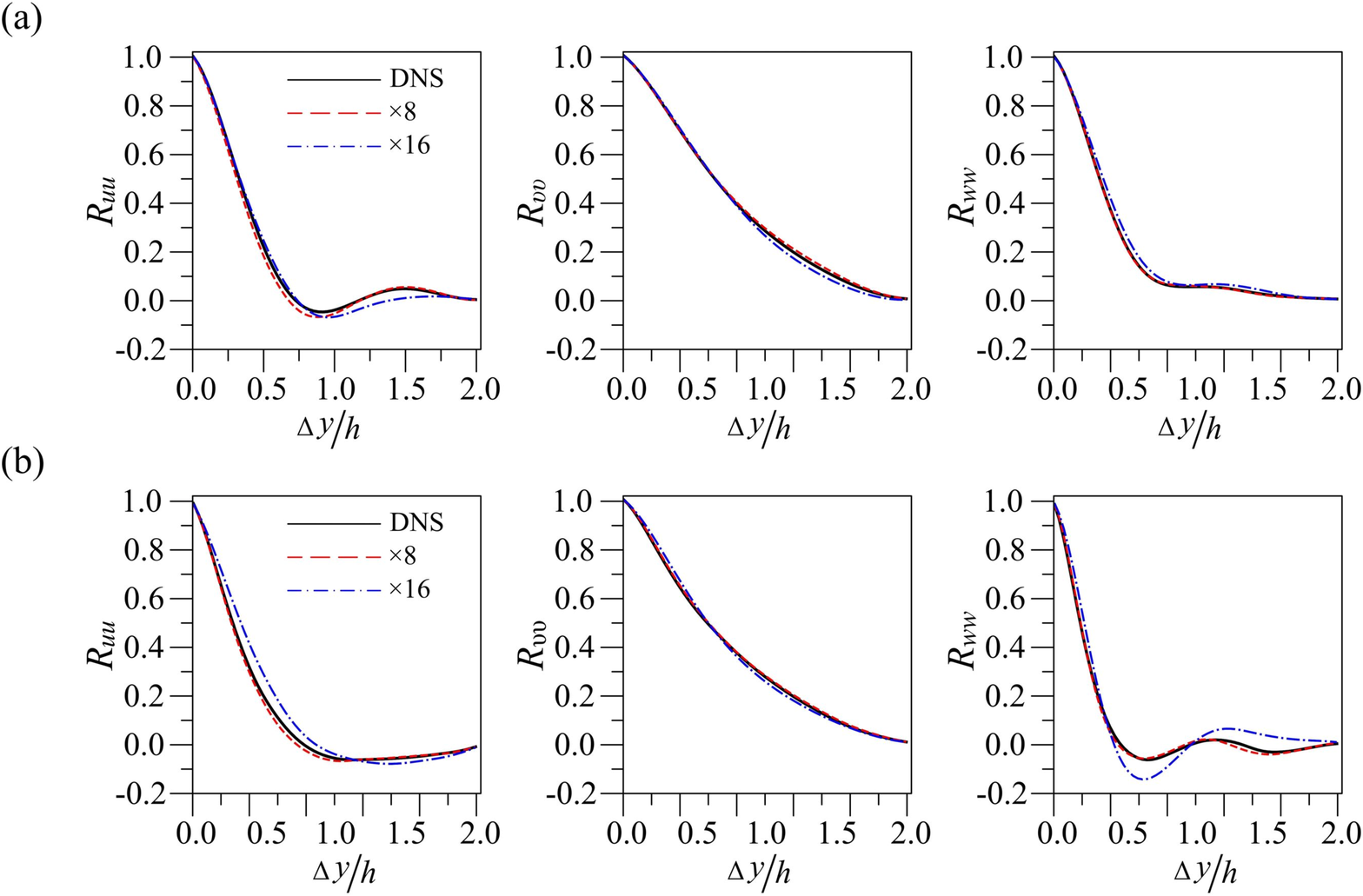}
\caption[]{Two-point correlations of the reconstructed velocity components at (a) $Re_\tau = 400$ and (b) $Re_\tau =600$} 
\label{fig:13-Rec2Correl-292-550}
\end{figure}

Figure~\ref{fig:13-Rec2Correl-292-550} shows $R_{ii} (\Delta y )$ of the reconstructed velocity components. The correlations reveal a good match with the DNS results, with no noticeable reduction in accuracy for increasing Reynolds numbers. As expected, an offset is observed when an extremely low-resolution is used for two of the Reynolds numbers. \par

The results in this section demonstrate that the model is able to reconstruct high-resolution velocity fields at Reynolds numbers falling within the range of the training process with acceptable precision. Similar to the case in which the training Reynolds numbers are used, the reconstruction precision exhibit less sensitivity to increasing Reynolds number than it does to decreasing resolution level. \par

\subsection{Reconstruction error}

\begin{figure}
\centering 
\includegraphics[angle=0, trim=0 0 0 0, width=1\textwidth]{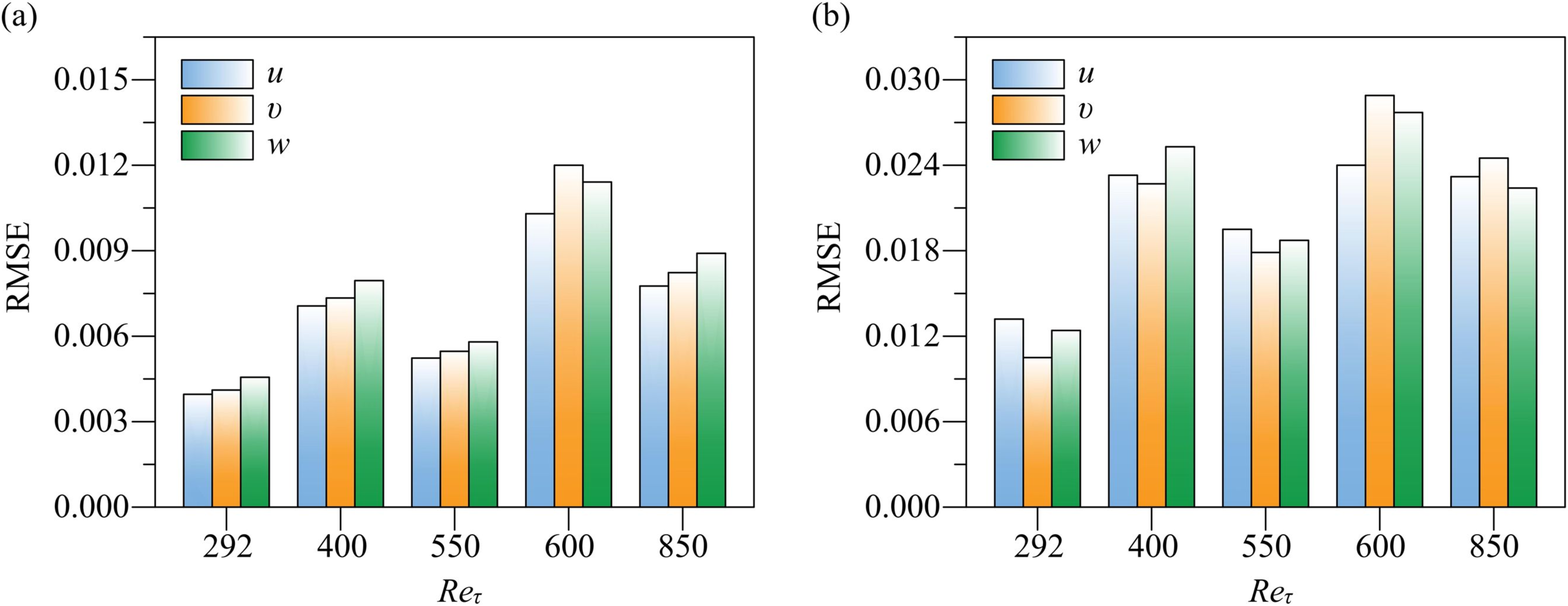}
\caption[]{Root mean square error of the reconstructed velocity components using two low-resolution levels: (a) $\times 8$ and (b) $\times 16$} 
\label{fig:14-RecRMSE}
\end{figure}

To further explore the performance of the model in terms of its reconstruction accuracy, the reconstruction error is statistically analysed using root mean square error (RMSE), as reported in Figure~\ref{fig:14-RecRMSE}. As observed in Figure~\ref{fig:14-RecRMSE}(a) and (b), the RMSE values for the case in which the training Reynolds numbers are used are proportional to the increase in the Reynolds number. As expected, the RMSE values tend to increase with the decrease of the resolution level due to the lack of information about the velocity fields when extremely low-resolution data are used. It can also be observed that when the reconstructed velocity fields are represented by Reynolds numbers not used in the training process, the error values rapidly increase and further increase with decreasing resolution level. \par

Hence, the RMSE can be represented as:

\begin{equation}\label{eqn:eq9}
{\rm RMSE} = {\rm RMSE} _{\rm Reynolds~number} + {\rm RMSE} _{\rm interpolation} + {\rm RMSE} _{\rm resolution}, 
\end{equation}

\noindent where ${\rm RMSE_{Reynolds~number}}$ is the error due to the increase of $Re_\tau$, ${\rm RMSE_{interpolation}}$ is the error due to the interpolation between the Reynolds numbers used in the training process to obtain the velocity fields at the Reynolds numbers falling within the training range, i.e. $Re_\tau$ = 292 \textendash 850 and ${\rm RMSE_{resolution}}$ is the error representing the effect of the resolution level. Based on this decomposition of the RMSE, we can see that the combined effect of ${\rm RMSE_{interpolation}}$ and ${\rm RMSE_{resolution}}$ can increase the error values by a factor of three. Nevertheless, since the maximum RMSE in this study is on the order of $10^{-2}$, the MS-ESRGAN performance is entirely acceptable when compared with the results obtained by Deng {\it et al.} \cite{Dengetal2019}. These results further indicate that MS-ESRGAN has remarkable potential for reconstructing high-resolution turbulent velocity fields from extremely low-resolution data at a wide range of Reynolds numbers not necessarily used in the training process.

\section{Conclusions}
This study presented a deep learning-based method to reconstruct high-resolution turbulent velocity fields from extremely low-resolution data at various Reynolds numbers. In particular, MS-ESRGAN was implemented to map the low-resolution velocity fields to the high-resolution ones. Velocity data of turbulent channel flow with large longitudinal ribs obtained from DNS were then used to validate the model. Three different Reynolds numbers were used in the training process. \par

First, the capability of the model to reconstruct high-resolution velocity fields at the Reynolds numbers used in training based on two different low-resolution levels was examined. The results revealed that the model could successfully reproduce the high-resolution instantaneous velocity fields with commendable accuracy, even for the lowest-resolution level used in this study. Although the PDF, $R_{ii} (\Delta y)$ and RMS profiles did exhibit a slight deviation when the lowest-resolution level was used, the results still had acceptable accuracy. \par

The ability of the model to reconstruct velocity fields at Reynolds numbers that were not used in the training process was then examined using two different Reynolds numbers that fell within the training Reynolds numbers range. Here, both the instantaneous velocity fields and statistical analysis results matched the DNS data well, indicating that the model has great potential to interpolate between velocity fields that lie within the training Reynolds number range. \par

The maximum reconstruction error also exhibited acceptable values, indicating MS-ESRGAN has remarkable ability to map extremely low-resolution velocity fields at a wide range of Reynolds numbers to high-resolution ones with commendable accuracy. \par

The results from this study indicate that GAN-based model, such as MS-SRGAN, has the ability to map limited spatial turbulence data to high-fidelity data using flow fields information represented by a few specified Reynolds numbers that determine the minimum, intermediate and maximum values of the interpolation. \par

\begin{acknowledgments}
This work was supported by 'Human Resources Program in Energy Technology' of the Korea Institute of Energy Technology Evaluation and Planning (KETEP), granted financial resource from the Ministry of Trade, Industry \& Energy, Republic of Korea (no. 20214000000140). In addition, this work was supported by the National Research Foundation of Korea (NRF) grant funded by the Korea government (MSIP) (no. 2019R1I1A3A01058576). 
\end{acknowledgments}

\section*{Data Availability}
The data that supports the findings of this study are available within this article.

\appendix

\nocite{*}
\bibliography{my-bib}

\end{document}